# The Evolution of Hyperedge Cardinalities and Bose-Einstein Condensation in Hypernetworks

Jin-Li Guo[1*], Qi Suo[1], Ai-Zhong Shen[1], Jeffrey Forrest[2]

**Abstract:** To depict the complex relationship among nodes and the evolving process of a complex system, a Bose-Einstein hypernetwork is proposed in this paper. Based on two basic evolutionary mechanisms, growth and preference jumping, the distribution of hyperedge cardinalities is studied. The Poisson process theory is used to describe the arrival process of new node batches. And, by using the Poisson process theory and a continuity technique, the hypernetwork is analyzed and the characteristic equation of hyperedge cardinalities is obtained. Additionally, an analytical expression for the stationary average hyperedge cardinality distribution is derived by employing the characteristic equation, from which Bose-Einstein condensation in the hypernetwork is obtained. The theoretical analyses in this paper agree with the conducted numerical simulations. This is the first study on the hyperedge cardinality in hypernetworks, where Bose-Einstein condensation can be regarded as a special case of hypernetworks. Moreover, a condensation degree is also discussed with which Bose-Einstein condensation can be classified.

As an effective tool to characterize complex systems, the studies of complex networks have attracted much attention in the past decade. This flurry of research was triggered by two seminal papers, that by Watts and Strogatz on small-world networks[1], and that by Barabási and Albert on scale-free networks[2]. Meanwhile, the massive analysis of complex networks has become a frontier in various scientific fields, such as physics, biology, computer science, sociology, economics and others. In a complex network, nodes correspond to different individuals involved in the study, while edges between nodes represent relationships between the connected nodes in the actual system. In complex networks, each edge connects only two nodes. However, relationships among the objects of each complex real life system tend to be more complex than those that can be described with simple pairwise relations, and can be readily described as a hypernetwork. In real world, complex systems, such as multi-machine systems[3], transportation systems[4], and research cooperation networks[5], can be more adequately described by using the concept of hypernetworks.

Recently, the concept of hypernetworks[6] has attracted much attention in the scientific community. Hyperlink prediction in hypernetworks using latent social features was studied[7]. Kim et al. predicted the clinical outcome of a cancer treatment by using evolving hypergraphs[8].

*[1]Business School, University of Shanghai for Science and Technology, Shanghai 200093, PR China. [2]School of Business, Slippery Rock University, Slippery Rock, PA 16057, USA. Correspondence should be addressed to J. L. G. (email: phd5816@163.com).

Komarov and Pikovsky reported on finite-sized-induced transitions to synchrony in a population of phase oscillators coupled via nonlinear mean field, which is microscopically equivalent to a hypernetwork organization of interactions[9].

However, few studies in the literature focus on dynamic evolution models of hypernetworks. Although the content of Wikipedia was described with a hypernetwork model[10], it represents just a special case of the general evolution model of complex networks in Ref.11. In addition, Hu *et al.*[5] proposed a hypernetwork model for scientific cooperation. Zhang and Liu[12] proposed a preferential attachment mechanism hypernetwork based on the users' background knowledge, objects, and labels. Wang *et al.*[13] offered a hypernetwork dynamic evolution model with growth and preferential attachment mechanisms, where at each time step some new nodes are added and connected to an old node through a hyperedge. Hu *et al.*[14] developed another hypernetwork dynamic evolution model that is similar to that of Wang's[13], where at each time step a new node is added and connected to some old nodes with a hyperedge. Although several hyperedges were added at each time step in Ref.15, the growth was the same as that studied in Hu's[14]. Consequently, Guo and Zhu[16] developed a hypernetwork evolution model that unified the hypernetwork models introduced in Refs. 13-15 and the BA model. Different from the preferential attachment mechanism as mentioned in the afore-mentioned literatures, Guo and Suo[17] proposed another model and introduced two other factors, such as node's own fitness and competitiveness, to underlie the mechanism of preferential attachment.

The common feature of all the hypernetwork models reviewed above is that node degree distribution is an extension of the concept of degree distributions in complex networks. Nevertheless, the cardinality of a hyperedge is also an important parameter. In the study of hypernetworks, one of the tasks is to depict the topological characteristics with the hyperedge cardinality.

In addition to the above researches, models of network dynamics based on quantum statistics have also been well studied. Gachechiladze *et al.*[18] studied the nonlocal properties of quantum hypergraph states. Bianconi *et al.*[19] proposed the concept of quantum geometric networks, which has many properties common to those of complex networks. Quantum geometric networks can be distinguished from Fermi-Dirac networks and Bose-Einstein networks that obey respectively the Fermi-Dirac and Bose-Einstein statistics. Kulvelis *et al.*[20] studied single-particle quantum transport on parametrized complex networks. Bianconi and Barabási[21] tried to map an equilibrium Bose gas into a complex network and found the emergence of Bose-Einstein condensation in such evolving networks, which is indeed a pioneering work on the subject matter. Other studies include bosonic networks[22], fermionic networks[23,24], and Bose-Einstein condensation in the Apollonian complex network[25]. Recently, Bianconi[26] constructed a multiplex network which is described by coupled Bose-Einstein and Fermi-Dirac quantum statistics. They extended the definition of entanglement entropy of multiplex structures.

However, in Ref. 21 each same energy level was regarded as a node, which resulted in such a consequence as that all particles at different energy levels shrink into a node. In Ref.26 the nodes in each layer were the same. The structural properties, including the degree distribution in different layers and different types of correlations have been obtained. Contrary to these works, the originality of our present work is to regard particles as nodes and energy levels as hyperedges, on which an evolving hypernetwork model is developed and its essential properties are studied. On one hand, nodes in different hyperedges can be different from each other. On the other hand, the cardinalities of hyperedges are dynamic changing as time goes on. One purpose of the paper is to obtain the stationary average hyperedge cardinality distribution. Furthermore, our model is able to capture the phenomenon of Bose-Einstein condensation in the evolution of hypernetworks.

The rest of the paper is organized as follows. In next section, we introduce respectively the concepts of hypergraphs and hypernetworks. After that, we propose Bose-Einstein hypernetwork model. That is followed by theoretical analysis and numerical simulations of our model. At the end, our presentation is concluded with some conclusion remarks.

**The Related Concepts**

The generalization of the concept of complex networks can be categorized into those of network-based supernetworks and hypergraph-based hypernetworks.

A supernetwork is a 'network of networks'. This concept was first proposed by Denning in 1985, while it was clearly defined by Nagurney[27]. In supernetworks, there are large scale and complex connections, resulting in many networks mingled with each other. Such networks are also called 'multilayer networks'[28,29] or 'multilevel networks'[30]. Usually they possess 'multi-stranded' relationships and are formed by layers. Each layer can be seen as a graph, and interconnections are existed between nodes of different layers. Such networks constitute a natural environment to describe systems interconnected through different categories of connections. For further information, please consult with review articles[28,6].

Another concept is that of hypergraph-based hypernetworks. Each edge in a hypergraph, known as hyperdeges, contains arbitrary number of nodes[31]. The extension from edge to hyperededge, make it possible to relate groups of more than two nodes. Meanwhile, the network structure is simple and clear. A complex system represented by a hypergraph will be referred to as a hypernetwork[32]. The following is the mathematical definition of hypergraphs and hypernetworks. The concept of hypergraphs generalizes that of graphs by allowing for edges of higher cardinality. Formally, we define a hypergraph as a binary $H = (V, E^h)$, which is also denoted as $(V, E^h)$ or $H$, where $V = \{v_1, v_2, \cdots, v_n\}$ ($|V|$ denotes the cardinality of the set $V$) and $E^h = \{E_1, E_2, \cdots, E_{|E^h|}\}$ ($|E^h|$ denotes the cardinality of the set $E^h$, $E_i = \{v_{i1}, v_{i2}, \cdots, v_{ik}\}$ ($v_{ij} \in V, j = 1,2,\cdots,k$)) are the sets of nodes and hyperedges, respectively. While for graphs edges connect only two nodes, each hyperedge can connect more than two nodes; to this end, examples of hypergraphs are depicted in Figure1. Two nodes are said to be adjoined if they belong to the same hyperedge. Two hyperedges are said to be adjoined if their intersection set is not empty. $H$ is said to be a finite hypergraph if both $|V|$ and $|E^h|$ are finite. A hypergraph $H = (V, E^h)$ is a k-uniform hypergraph if $|E_i| = k$, for $i = 1,2,\cdots,|E^h|$. An example of four-uniform hypergraph is depicted in Figure1(a).

With the definitions above, we now establish the concept of hypernetworks. Assuming that $\Omega = \{(V, E^h) | (V, E^h) \text{ is a finite hypergraph}\}$ and $G^h$ is a mapping from $[0, +\infty)$ to $\Omega$, where $G^h(t) = (V(t), E^h(t))$ is a finite hypergraph for any given $t \geq 0$. Here the indicator $t$ is often interpreted as time. A hypernetwork is a set of hypergraphs. The degree (or hyperdegree) of node $v_i$ is defined as the number of hyperedges containing the node. The cardinality of a hyperedge $E_i$ is defined as the number of nodes contained in the hyperedge.

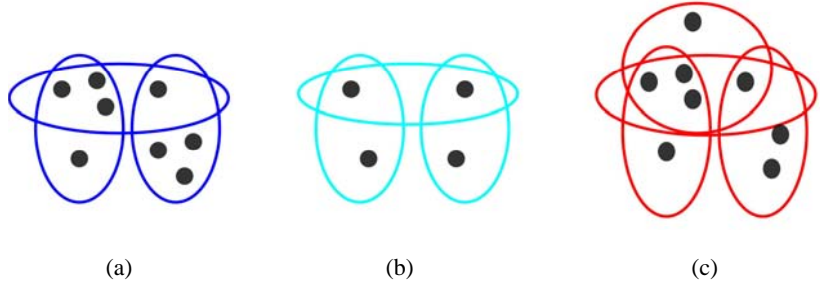

(a)                            (b)                            (c)

**Figure1.** Examples of hypergraphs. (a) A four-uniform hypergraph. (b) An ordinary graph. When each edge in a hypergraph contains only two nodes, the hypergraph degenerates into an ordinary graph. (c) A non-uniform hypergraph.

## Model Description

*Hypernetwork model*

In the real-life world, hyperedge growth and hyperedge preferential attachment are the bases of the evolution mechanism. The generation algorithm in Ref.13 can be described as follows. (1) The hypernetwork starts with $m_0$ nodes and a hyperedge which concludes all these $m_0$ nodes. At each time step, $m_1$ nodes are added to the system, and a new hyperedge is constructed by connecting these new nodes and an existing old nodes. (2) The probability $\prod(d_H(i))$ that an old node $i$ is selected by the new hyperedge depends on the hyperdegree $d_H(i)$ of node $i$:

$$\prod d_H(i) = \frac{d_H(i)}{\sum_j d_H(j)}. \tag{1}$$

When analyzing this model, Ref. 13 assumed that node batches are discretely added to the system at time $t\,(=0,1,2,3,\cdots)$. For other evolving models in hypernetworks, please consult with Refs.14,15.

*Bose-Einstein hypernetwork*

Most of the afore-mentioned models are *k*-uniform hypernetworks. Results of theoretical analysis show that hyperdegree distributions exhibit the scale-free property. Different from the models described previously, we will construct a non-uniform hypernetwork. In particular, the evolving mechanism reflects the common feature of competitions among hyperedges, resulting in the evolution of the cardinality of hyperedges. Here we show that our model can be mapped into an equilibrium Bose gas for treating particles as nodes while considering different energy levels as hyperedges.

For the growth mechanism, it is often assumed that the nodes are added to the system at equal-lengthtime intervals, and the arrival of nodes follows a uniform distribution. The continuum assumption of discrete problems is the precondition for analyzing node degrees by employing differential equations[2]. Here we assume that the process of node arrivals follows a Poisson process to better describe the arrival patterns in realistic systems, and hence this assumption allows us a more rigorous analysis of the model.

For the preference mechanism, how can we reflect the state of competition among hyperedges? It is clear that a hyperedge with bigger cardinality has more probability to be selected. Another dimension is the energy level of a hyperedge, which closely relates to the hyperedge's competitiveness. These two parameters jointly underlie the evolution process. Our hypernetwork model satisfies the following two steps.

(1) Growth: The arrival process of new node batches is a Poisson process $N(t)$ with a constant rate $\lambda$. At time $t$, when a batch of new nodes arrives at the network, a positive integer $\varsigma_{N(t)}$ is chosen from a given distribution $g(\varsigma)$, accounting for the number of new nodes. The new $\varsigma_{N(t)}$ nodes are encircled by a new hyperedge $E_{N(t)}$, while the energy $\varepsilon_{N(t)} > 0$ of hyperedge is drawn from a given distribution $\rho(\varepsilon)$. And each new node is assigned to a state.

(2) Preferential jump: At time $t$, a hyperedge is randomly chosen from the hypernetwork. And a randomly chosen node that belongs to this hyperedge jumps into another hyperedge. The probability $W$ that the chosen node jumping into hyperedge $j$ depends on the cardinality $h_j$ of the hyperedge $j$ and on the energy level $\varepsilon_j$ of the hyperedge $j$ such that

$$W(h_j) = \frac{e^{-\beta\varepsilon_j} h_j}{\sum_j e^{-\beta\varepsilon_j} h_j}, \qquad (2)$$

where $\beta = \frac{1}{T}$, $T$ is temperature, $m = \int \varsigma g(\varsigma) d\varsigma < +\infty$ is the expected value of the initial cardinality of all hyperedges. .

In some cases, the fitness $\eta_j$ of hyperedge $j$ is determined by its energy level. The relationship between the fitness and energy level of the hyperedge $j$ is given as follows:

$$\eta_j = e^{-\beta \varepsilon_j}. \qquad (3)$$

Here we interpret how the hypernetwork model above corresponds to a realistic system. Take online shopping as an example; its features can be depicted by the model above. In electronic commerce, with new products and new customers joining into the network, the sales network is in a constant state of growth. One obvious characteristic of online purchasing behaviors is that customers tend to purchase products associated with good quality and higher sales. This reflects the mechanism of preferential purchase. Customers have preferences for higher quality items. However, it is hard for them to differentiate products in terms of their qualities. In many cases, historical sales data is an effective tool for customers to make their decisions based on the assumption that better products are accompanied by higher sales. Treating customers as nodes while considering products as hyperedges, the fitness and the cardinality of hyperedges represent respectively the competitiveness and the sales history of products. At the beginning, purchasing behaviors are scattered. With the evolution of the network, products with higher competitiveness and sales will attract more customers, thus resulting in preferential jumpings.

A schematic illustration of the dynamical rules for building a Bose-Einstein hypernetwork is shown in Figure2.

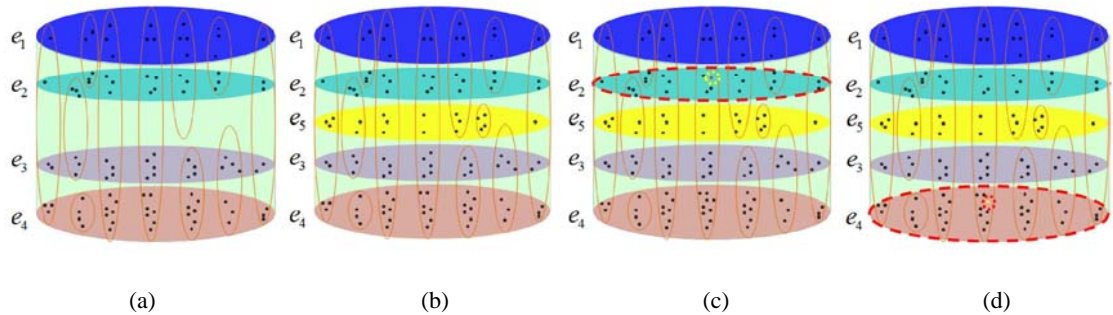

(a)　　　　　　　　(b)　　　　　　　　(c)　　　　　　　　(d)

**Figure2.** Schematic illustration of the process at each time step of a Bose-Einstein hypernetwork. (a) At time step $t$, there are four energy hyperedges $e_1 \sim e_4$ described by closed curves, which contain different number of nodes respectively. Each node is assigned a state. And nodes with the same state are encircled by a red ellipse representing a state hyperedge. (b) At time step $t+1$, a batch of new nodes encircled by a new energy hyperedge $e_5$ arrives at the network. (c) Select an energy hyperedge randomly ($e_2$ is selected and shown as a red hollow ellipse) from the existing network. Then select a node randomly in $e_2$ (shown as the red one encircled by a yellow hollow ellipse). (d) The selected yellow node jumps from the original energy hyperedge $e_2$ to another energy hyperedge $e_4$ according to the preferential mechanism. The quantum state of the node remains unchanged while jumping.

**Model Analyses**

Here we focus on the dynamics of hyperedge cardinality. Firstly, we write down the rate equation for the distribution of hyperedge cardinalities. Then the theoretical results are given based on the Poisson process theory and a continuity technique[21].

Let $N(t) = \{$the number of node batches in the network at time $t\}$. The arrival process of node batches is a Poisson process with constant rate $\lambda$. According to the Poisson process theory, $E[N(t)] = \lambda t$. Let $h_j(t)$ represent the cardinality of the $j$ th-hyperedge at time $t$. Assume that $h_j(t)$ is a continuous real variable. By the assumption of continuity, we know that $h_j(t)$ satisfies the following dynamic equation

$$\frac{\partial h_j(t)}{\partial t} = \lambda \frac{e^{-\beta \varepsilon_j} h_j}{\sum_j e^{-\beta \varepsilon_j} h_j} - \frac{1}{t}. \tag{4}$$

The first term in the right of Eq. (4) corresponds to preferential attachment of a hyperedge that is selected by a node. And the second term corresponds to the randomly selection of a node to jump out of the current hyperedge.

Let

$$x = \lim_{t \to \infty} \frac{1}{\lambda t} \sum_j e^{-\beta \varepsilon_j} h_j(t).$$

For sufficiently large $t$, we have

$$\sum_j e^{-\beta \varepsilon_j} h_j(t) = \lambda t x. \tag{5}$$

Substituting Eq.(5) into Eq.(4), we have

$$\frac{\partial h_j(t)}{\partial t} = \frac{e^{-\beta \varepsilon_j} h_j(t)}{xt} - \frac{1}{t}. \tag{6}$$

Because of the initial condition that each hyperedge $j$ satisfies $h_j(t_j) = \varsigma_j$, the solution of Eq.(6) is

$$h_j(t,\varsigma) = (\varsigma_j - \frac{x}{e^{-\beta\varepsilon_j}})(\frac{t}{t_j})^{\frac{e^{-\beta\varepsilon_j}}{x}} + \frac{x}{e^{-\beta\varepsilon_j}}, \tag{7}$$

where $x$ is the positive solution of the following equation

$$(m-1)\int \frac{e^{-\beta\varepsilon}}{x - e^{-\beta\varepsilon}} \rho(\varepsilon)d\varepsilon = 1. \tag{8}$$

Eq.(8) is called a characteristic equation of the hyperedge cardinality of the Bose-Einstein hypernetwork.

According to the Possion process theory, the arrival time $t_j$ of node batches obeys Gamma distribution having parameters $(i,\lambda)$, thus

$$P\{h_j(t,\varsigma,\varepsilon) \geq k\} = 1 - e^{-\lambda t(\frac{\varsigma_j e^{-\beta\varepsilon_j} - x}{ke^{-\beta\varepsilon_j} - x})^{\frac{x}{e^{-\beta\varepsilon_j}}}} \sum_{l=0}^{j-1} \frac{1}{l!}(\lambda t(\frac{\varsigma_j e^{-\beta\varepsilon_j} - x}{ke^{-\beta\varepsilon_j} - x})^{\frac{x}{e^{-\beta\varepsilon_j}}})^l. \tag{9}$$

From Eq.(9), we obtain the stationary average hyperedge cardinality distribution of the Bose-Einstein hypernetwork as follows

$$P(k) = \int g(\varsigma)d\varsigma \int \frac{x}{\varsigma e^{-\beta\varepsilon} - x}(\frac{\varsigma e^{-\beta\varepsilon} - x}{ke^{-\beta\varepsilon} - x})^{xe^{\beta\varepsilon}+1} \rho(\varepsilon)d\varepsilon, \tag{10}$$

where $x$ is the positive solution of Eq.(8).

For simplicity, for given $\varsigma = m$, the stationary average hyperedge cardinality distribution of the Bose-Einstein hypernetwork is as follows

$$P(k) = \int \frac{\theta}{me^{-\beta\varepsilon} - \theta}(\frac{me^{-\beta\varepsilon} - \theta}{ke^{-\beta\varepsilon} - \theta})^{\theta e^{\beta\varepsilon}+1} \rho(\varepsilon)d\varepsilon, \tag{11}$$

where $\theta$ is the positive solution of Eq. (8).

When the energy level $\varepsilon_{N(t)}$ is taken from the uniform distributions over $[0,1]$, the hyperedge cardinality distribution is

$$P(k) = \int_0^1 \frac{\theta}{me^{-\beta\varepsilon} - \theta}(\frac{me^{-\beta\varepsilon} - \theta}{ke^{-\beta\varepsilon} - \theta})^{\theta e^{\beta\varepsilon}+1} d\varepsilon, \tag{12}$$

where

$$\theta = \frac{\exp(\beta/(m-1)) - \exp(-\beta)}{\exp(\beta/(m-1)) - 1}. \tag{13}$$

If $k \gg \theta$, from Eq. (12) we obtain

$$P(k) = \int_0^1 \theta(m - \theta e^{\beta\varepsilon})^{\theta e^{\beta\varepsilon}} \frac{1}{k^{\theta e^{\beta\varepsilon}+1}} d\varepsilon. \tag{14}$$

Using Eq. (14) we obtain,

$$P(k) \propto \frac{k^{-\theta} - k^{-e^{\theta\beta}}}{k \ln k}. \tag{15}$$

*i.e.* the hyperedge cardinality distribution follows a generalized power law, with an inverse logarithmic correction[33].

Numerical simulations for the distributions of hyperedge cardinalities are given. The simulations are performed with the scale of $N = 100000$ (the total number of nodes), and each simulation result is obtained by averaging over 30 independent runs. The simulation results are shown in Figures3-4 in double-logarithmic axis. From the evolution mechanism of the model, we know that the cardinalities and energy levels of hyperedges jointly determine the evolution. Thus the ability for hyperedges to compete for nodes is not the same from one hyperedge to another. Nodes tend to jump to the most attractive hyperedges, and these hyperedges thus acquire more and more nodes over time. And this process results in that a tiny fraction of the hyperedges will acquire respectively good numbers of nodes. As the figures show, the theoretical prediction result which is obtained from Eq.(15) is consistent with the tail of the distributions of hyperedge cardinalities in simulations.

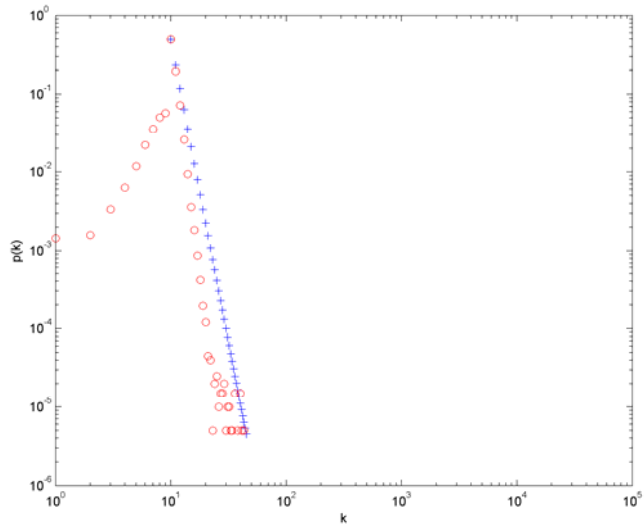

**Figure3.** The number of nodes is equal to 100000, the number of new nodes is equal to 10, β =1, the energy level follows a uniform distribution on [0, 1]. O denotes the simulation result, and + the theoretical prediction.

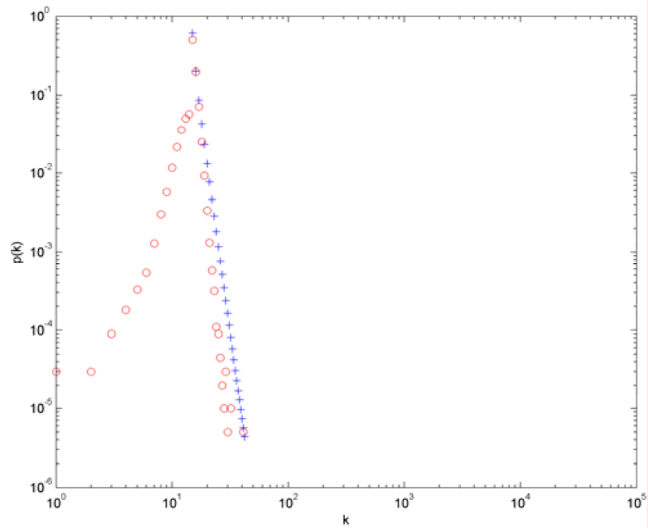

**Figure4.** The number of nodes is equal to 100000, the number of new nodes is equal to 15, β =1, the energy level follows a uniform distribution on [0, 1]. O denotes the simulation result, and + the theoretical prediction.

*Bose-Einstein condensation*

Assume that $M_{\varepsilon=0}$ represents the number of nodes on the energy level where $\varepsilon=0$, then we have

$$\frac{1}{N(t)}M_{\varepsilon=0}=1-(m-1)\int\frac{e^{-\beta\varepsilon}}{x-e^{-\beta\varepsilon}}\rho(\varepsilon)d\varepsilon. \qquad (16)$$

If $\frac{1}{mN(t)}M_{\varepsilon=0}\geq\alpha\ (0<\alpha\leq 1)$, then the nodes condense on the energy level of $\varepsilon=0$ where $\alpha$ is called a condensation degree of the hypernetwork. The condition the condensation degree $\alpha$ on the energy level where $\varepsilon=0$ satisfies is given as follows

$$\int\frac{e^{-\beta\varepsilon}}{x-e^{-\beta\varepsilon}}\rho(\varepsilon)d\varepsilon\leq\frac{1-\alpha m}{m-1}. \qquad (17)$$

If $\alpha=1$, the hypernetwork almost completely condenses on the energy level of $\varepsilon=0$.

Regarding particles as nodes, Bose-Einstein condensation can be described by the model above. According to the condensation degree, Bose-Einstein condensations can be classified.

The particles of a Bose-Einstein condensation model follow the stationary average cardinality distribution Eq.(10) at each energy level. By introducing the concept of chemical potential $\mu$, we let $x = e^{-\beta\mu}$,

$$I(\beta,\mu) = \int_0^{+\infty} \frac{1}{e^{\beta(\varepsilon-\mu)}-1}\rho(\varepsilon)d\varepsilon. \quad (18)$$

Since for any given $\varepsilon > 0$, $\frac{1}{e^{\beta(\varepsilon-\mu)}-1} \geq 0$, we have $\mu \leq 0$. That is, the chemical potential is nonpositive.

The maximum of $I(\beta,\mu)$ is obtained when $\mu = 0$, for given $\beta$, $m$, and $\rho(\varepsilon)$; thus we have

$$I(\beta,0) = \int_0^{+\infty} \frac{1}{e^{\beta\varepsilon}-1}\rho(\varepsilon)d\varepsilon \leq \frac{1-\alpha m}{m-1}. \quad (19)$$

The condensation degree is $\alpha$ on the lowest energy level.

From Eq.(17), it follows that Bose-Einstein condensation appears when Eq. (8) has no solution, at which point Eqs. (7) and (8) break down. The absence of a solution indicates that almost all hyperedges have only a few of nodes, while some "gel" hyperedges have the rest of the nodes of the hypernetwork. This end seems to be a well-known signature of Bose-Einstein condensation.

## Conclusions

By taking into account the fact that the concept of hypernetworks is more general as hypernetwroks allow for the dynamics of hyperedge cardinality and node degree, we propose Bose-Einstein hypernetwork evolution model by combining the growth and preferential jump mechanisms to investigate hyperedge cardinalities of the hypernetwork structure. We obtain the distribution of hyperedge cardinalities by using theoretical analysis and numerical simulations. Our simulation results are in good agreement with theoretical conclusions. Specially, when treating particles as nodes while considering different energy levels as hyperedges, the Bose-Einstein condensation model can be regarded as a special case of our model. Furthermore, we establish the condensation condition of the hypernetwork on the zero-energy level. The solid results established in this paper lead us to believe that the concept of hypernetworks can be used as a new tool for the study of statistical physics.

Presently, the research on the topological characteristics and evolution mechanism of hypernetworks are just started. Although we have obtained some essential theoretical result of the model, the corresponding empirical studies are still absent. Such empirical studies, as a complement to the model, can further enrich the current research. Besides hyperdegrees and cardinalities, there are other important parameters, such as clustering coefficient, assortativity that have been discussed in the study of complex networks. The definitions and analyses of such parameters in hypernetworks need to be explored. Furthermore, there is a real need to understand whether or not hypernetworks pose common topological features in a self-organized way. Followup studies could focus on the evolutionary dynamics of hypernetwork structures and dynamical processes that occur over hypernetworks.

## Acknowledgements

This work is supported by National Natural Science Foundation of China under the grant no. 71571119, Science Foundation of Ministry of Education of China under the grant no. 16YJC870012. The authors are grateful to the anonymous referees for their helpful comments that improved this paper.

## Author information

Affiliations

Business School, University of Shanghai for Science and Technology, Shanghai, 200093, China

Jin-Li Guo Qi Suo Ai-Zhong Shen

School of Business, Slippery Rock University, Slippery Rock, PA, 16057, USA

Jeffrey Forrest

## Author Contributions

J.-L. G. designed the research, proposed the models, performed the data analysis and wrote the paper; Q. S. helped with the interpretation of the theoretical analysis, as well as revised the manuscript. A. Z. S. revised the manuscript; J. F. participated in the English revision. All authors reviewed the results and approved the final version of the manuscript.

## Competing interests

The author declares no competing financial interests.

## Corresponding author

Correspondence to Jin-Li Guo.